\documentclass[a4paper, 10pt]{report}

\usepackage[cp1250]{inputenc}
\usepackage[T1]{fontenc}
\usepackage{amsfonts}
\usepackage{graphicx}
\usepackage{amsmath}
\usepackage{caption}

\usepackage{amssymb}

\usepackage{fancyhdr}
\usepackage{bibtopic}
\usepackage{color}
\usepackage{ulem}
\usepackage[toc,page]{appendix}
\usepackage{setspace}
\usepackage{cite}
\usepackage{xcolor}

\AtBeginDocument{}
\usepackage{etoolbox}
\patchcmd{\thebibliography}{\chapter*}{\section*}{}{}

\makeatletter
\renewcommand{\thesection}{%
  \ifnum\c@chapter<1 \@arabic\c@section
  \else \thechapter.\@arabic\c@section
  \fi
}
\makeatother

\patchcmd{\tableofcontents}{\chapter*}{\section*}{}{}

\numberwithin{equation}{section}

\let\OLDthebibliography\thebibliography
\renewcommand\thebibliography[1]{
  \OLDthebibliography{#1}
  \setlength{\parskip}{0pt}
  \setlength{\itemsep}{3.5pt plus 1ex}
}

\usepackage[linktocpage]{hyperref}
\definecolor{darkred}{rgb}{0.5,0,0}
\definecolor{darkpurple}{rgb}{0.5,0,0.5}
\definecolor{darkblue}{rgb}{0,0,0.5}
\hypersetup{ colorlinks,
linkcolor=darkblue,
filecolor=darkpurple,
urlcolor=darkred,
citecolor=darkpurple }
 
\usepackage{soul}

\begin{document}

\allowdisplaybreaks
\setlength{\abovedisplayskip}{3.5pt}
\setlength{\belowdisplayskip}{3.5pt}
\abovedisplayshortskip
\belowdisplayshortskip

{\setstretch{1.0}

{\LARGE \bf \centerline{
Solid matter with zero shear modulus
}}
{\LARGE \bf \centerline{
in flat universe
}}

\vskip 1cm
\begin{center}
{Peter M\'esz\'aros\footnote{e-mail address: peter.meszaros@fmph.uniba.sk}}

\vskip 2mm {\it Department of Theoretical Physics, Comenius
University, Bratislava, Slovakia}

\vskip 2mm \today 
\end{center}

\section*{Abstract}

For a perfect fluid, the quantity defined through mixed components of the stress-energy tensor $\widetilde{w}=(T_{i}^{\phantom{i}i}/3)/(-T_{0}^{\phantom{0}0})$ is independent on the choice of coordinates only for two values of the pressure to energy density ratio $w=p/\rho$: for radiation with $w=1/3$, and for dark energy with $w=-1$.
With other choices of $w$, the quantity $\widetilde{w}$ is coordinate dependent, and $\widetilde{w}=w$ only in the local rest frame of the fluid.
We show that the same is true also for solid matter with shear stress Lam\'e coefficient set to zero in a flat Friedmann--Lema\^itre--Robertson--Walker universe with perturbed metric as well as stress-energy tensor.
We call the two different solids with coordinate independent $\widetilde{w}$ radiation-like solid and dark energy-like solid,
and we restrict ourselves to these two special cases.
By analysing second order perturbations we discover two one parametric sets of such solid matter models containing special cases of radiation and dark energy as perfect fluids.
We also study equations for perturbations up to the second order for both sets of models.

\vskip 3mm \hspace{4mm}
\begin{minipage}[t]{0.8\textwidth}
\noindent\rule{12cm}{0.4pt}
\vspace{-9mm}
\tableofcontents
\noindent\rule{12cm}{0.4pt}
\end{minipage}
\vskip 6mm

}

\section{Introduction}\label{sec:1}

Theory of cosmological perturbations \cite{bardeen} is the key tool for comparing theoretical models with observations of the cosmic microwave background (CMB) radiation \cite{weinberg,cmbslow}. Thanks to this theory and modern observational data the values of cosmological parameters are known with high accuracy \cite{planck}. Properties of primordial perturbations serving as initial conditions for evolution of cosmological perturbations within the $\Lambda$-CDM model can be deduced from this comparison as well. This constrains the parameter space of models of cosmic inflation \cite{planck2,planck3}.

\vskip 1mm
From the time after the end of inflation and reheating to the time when nonlinear effects cause growth of the structure, the matter content of the universe can be considered as a multicomponent fluid with viscosity due to interaction between its components \cite{silk}. However, one can speculate on more peculiar forms of matter filling the universe, or rather one has to do so in order to study inflation or dark energy and dark matter sectors. The prominent example of matter generalizing the perfect fluid is scalar field which can drive the inflation \cite{brout,linde}, describe dark energy \cite{weinberg2}, dark matter \cite{sin}, or even both of them at the same time \cite{bento,scherrer}. Among more exotic matter components of the universe there may be also solid matter with elastic properties - i.e. matter with nonzero shear modulus. Similarly as the scalar field, the solid matter can be part of the dark matter sector \cite{bucher}, or it can drive the inflation \cite{gruzinov,endlich,akhshik,bartolo}.
A very recent and more systematic approach generalizing the solid matter in cosmology is in \cite{cabass}.

\vskip 1mm
Despite the framework of cosmological perturbations within the standard $\Lambda$-CDM model and its inflationary extensions being successful in fitting observational data, there are still several discrepancies \cite{schwarz}. The most important problems include tension between the local value of the Hubble constant and its value from CMB observations \cite{riess,riess2}, and anomalies of the CMB anisotropies for low multipole moments \cite{land,pinkwart}. Especially the latter problem must be studied with the use of the perturbation theory \cite{gordon,erickcek}. In this context a natural modification is based on considering anisotropy of the background. Bianchi type anisotropy does not produce strong anomaly \cite{pontzen}, but a more recent works studying different kinds of the background anisotropy \cite{chang}, which may be caused by cosmic strings \cite{yang,jazayeri} or domain walls \cite{jazayeri2,firouzjahi}, seem to be more promising. Another example of a recent attempt to explain this CMB anomaly is based on cosmological model with bounce \cite{agullo}. So far, these problems remain open from both observational and theoretical points of view.

\vskip 1mm
Here we focus on matter filling a universe with flat Friedmann--Lema\^itre--Robertson--Walker (FLRW) metric which shares some properties with both perfect fluid and solid matter. Up to the first order of the perturbation theory it behaves in the same way as a perfect fluid, but at the same time it obeys equation of state of a solid matter. This manifests in deviation from the perfect fluid-like behaviour in second and higher orders of the perturbation theory. Thus we have to use the theory of cosmological perturbations with nonlinear effects \cite{muchanov,abramo,acquaviva,noh,nakamura,malik} taken into account. We describe this matter in the framework of general relativistic elasticity \cite{carter}.
Our approach may be developed even further inspired by soft matter in condensed matter physics, see for instance a very recent work bringing this concept to cosmology \cite{saridakis}, but in this paper we focus on the special case of solid described above.
This approach can be used for describing also a perfect fluid, as long as its volume elements are not displaced too far from their initial positions. Fortunately, an almost homogeneous perfect fluid filling an almost homogeneous universe is such a case. In other words, a slightly perturbed fluid filling the universe can be considered as a special case of solid matter.

\vskip 1mm
Our main goal is to study physically interesting models of solid matter with zero shear modulus in a cosmological context.
In order to do so, we restrict ourselves to two special cases corresponding to the same pressure to energy density ratio as for radiation, $w=1/3$, and for dark energy, $w=-1$.
The first case describes our early universe at the beginning of the standard Friedmann expansion within the standard $\Lambda$-CDM model, while the second case approximates either far future of the universe or the cosmic inflation before the standard Friedmann expansion.

\vskip 1mm
Both radiation and dark energy satisfy another condition which may be imposed on the solid in order to restrict its parameter space even more.
In particular, a relation for the components of the mixed stress-energy tensor
\begin{eqnarray}
\label{eq:condition}
\widetilde{w}=\frac{T_i^{\phantom{i}i}/3}{-T_0^{\phantom{0}0}} = \textrm{const.},
\end{eqnarray}
with $T_i^{\phantom{i}i}$ denoting space part trace, holds in arbitrary coordinates.
However, in the local rest frame of the fluid $\widetilde{w}=w$ for any value of pressure to energy density ratio, not only for $w=1/3$ or $w=-1$.
For radiation, or photon gas, the relation (\ref{eq:condition}) is the consequence of scale invariance of the Maxwell theory.
Due to conformal invariance, the full trace of the mixed stress-energy tensor, which is invariant under space-time diffeomorphisms, is zero.
This property of the stress-energy tensor, $T_{\mu}^{\phantom{\mu}\mu}=T_0^{\phantom{0}0}+T_i^{\phantom{i}i}=0$ in arbitrary coordinates, implies that by passing to another coordinates $T_0^{\phantom{0}0}$ changes by a factor opposite to change in $T_i^{\phantom{i}i}$, leaving $\widetilde{w}$ in (\ref{eq:condition}) unchanged.
As for dark energy, it can be described through the cosmological constant, so that its stress-energy tensor must be proportional to the space-time metric.
The condition (\ref{eq:condition}) then gives $T_0^{\phantom{0}0}/T_i^{\phantom{i}i}=g_0^{\phantom{0}0}/g_i^{\phantom{i}i}=1/3$, or $\widetilde{w}=w=-1$ in arbitrary coordinates.

\vskip 1mm
Radiation and dark energy are the only two kinds of matter satisfying (\ref{eq:condition}) in arbitrary coordinates, while for other kinds of matter it holds only in their local rest frame.
A more technical approach to this fact, based on analysis of the stress-energy tensor for perfect fluids can be found in appendix \ref{app:a}.

\vskip 1mm
The bulk of this paper can be schematically divided into three main parts:
\begin{itemize}
\item In section \ref{sec:2} we introduce matter which departs from the perfect fluid only when second or higher order perturbations are taken into account. It is solid matter with shear modulus set to zero.
\item In section \ref{sec:3} we treat this matter pertubatively with flat FLRW space-time metric as the background focusing on radiation-like, $w-1/3$, and dark energy-like, $w=-1$, cases. We investigate consequences of the constraint on the mixed stress-energy tensor -- validity of (\ref{eq:condition}) in arbitrary coordinates -- which not only allows only two above mentioned values of $w$, but also leads to additional restrictions.
\item In sections \ref{sec:4} and \ref{sec:5} we present equations for up to second order scalar perturbations for the model introduced in section \ref{sec:2} in two special cases corresponding to constraints studied in section \ref{sec:3}.
\end{itemize}
Finally, in section \ref{sec:6} we summarize and discuss the main results.
Throughout the paper we use the signature of the space-time metric $(-,+,+,+)$, together with units in which $c=1$.

\section{Solid matter with zero shear modulus}\label{sec:2}

In order to describe a solid matter in general relativity one has to use map from body space of the solid to the space-time,
$\phi^I \mapsto x^i$,
where body coordinates $\phi^I$ label elements of the solid.
If the solid is relaxed, we may choose the space coordinates to coincide with the body coordinates,
$x^i = \delta^i_I \overline{\phi}^I$.
It is convenient to describe deformations of the solid by the body metric which is defined as a pull-back of the space-time metric,
\begin{eqnarray}
\label{eq:bodymetric}
B^{IJ} = g^{\mu\nu} \phi^{I}_{\phantom{I},\mu} \phi^{J}_{\phantom{J},\nu},
\end{eqnarray}
and in general the dependence of the solid matter energy density on its deformations can be encoded by the equation of state of the form $\rho = \rho \left( B^{IJ} \right)$.
Furthermore, after restricting to the case of solid with homogeneous and isotropic properties the equation of state must be invariant with respect to global translations and rotations in the body space,
\begin{eqnarray}
\label{eq:symmetry}
\phi^I \mapsto \phi^I + T^I + R^I_{\phantom{I}J} \phi^J, \quad T^I \in \mathbb{R}, \quad R^I_{\phantom{I}J} \in SO(3).
\end{eqnarray}
Translational invariance is satisfied due to partial derivatives in the definition of the body metric (\ref{eq:bodymetric}) automatically, and the rotational invariance can be obtained by parametrizing the equation of state by quantities which are invariant with respect to global rotations.

\vskip 1mm
There are at most three independent quantities invariant with respect to (\ref{eq:symmetry}).
They can be chosen as
\begin{eqnarray}
\label{eq:invariants}
& & \left[ \delta B \right] = \overline{B}_{IJ} \delta B^{IJ},
\quad \left[ \delta B^2 \right] = \overline{B}_{IJ} \delta B^{JK} \overline{B}_{KL} \delta B^{LI}, \\
& & \left[ \delta B^3 \right] = \overline{B}_{IJ} \delta B^{JK} \overline{B}_{KL} \delta B^{LM}\overline{B}_{MN} \delta B^{NI}, \nonumber
\end{eqnarray}
where  $\overline{B}_{IJ}$ are components of the body metric corresponding to the relaxed state, and $\delta B^{IJ} = B^{IJ} - \overline{B}^{IJ}$ is given by the difference between the actual body metric and its relaxed configuration.
The equation of state with desired properties then can be parametrized as
\begin{eqnarray}
\label{eq:eqstate1}
\rho = \overline{\rho} \sum_{a=0}^{\infty} \sum_{b=0}^{\infty} \sum_{c=0}^{\infty} C_{abc} \left[ \delta B \right]^a \left [ \delta B^2 \right]^b \left[ \delta B^3 \right]^c,
\end{eqnarray}
and the stress-energy tensor corresponding to such solid matter is
\begin{eqnarray}
\label{eq:stressenery}
T_{\mu\nu} & = & 2 \frac{\partial \rho}{\partial B^{IJ}} \phi^I_{\phantom{I},\mu} \phi^J_{\phantom{J},\nu} - \rho g_{\mu\nu} = \\
& = & \overline{\rho} \sum_{a=0}^{\infty} \sum_{b=0}^{\infty} \sum_{c=0}^{\infty} C_{abc} \left[ \delta B \right]^a \left [ \delta B^2 \right]^b \left[ \delta B^3 \right]^c \nonumber\\
& & \left[ 2 \left( a \frac{\Psi^{(0)}_{\mu\nu}}{\left[ \delta B \right]} + 2b \frac{\Psi^{(1)}_{\mu\nu}}{\left[ \delta B^2 \right]} + 3c \frac{\Psi^{(2)}_{\mu\nu}}{\left[ \delta B^3 \right]} \right) - g_{\mu\nu} \right], \nonumber
\end{eqnarray}
where
\begin{eqnarray}
\Psi^{(0)}_{\mu\nu} & = & \overline{B}_{IJ} \phi^I_{\phantom{I},\mu} \phi^J_{\phantom{J},\nu}, \\
\Psi^{(1)}_{\mu\nu} & = & \overline{B}_{IK} \delta B^{KL} \overline{B}_{LJ} \phi^I_{\phantom{I},\mu} \phi^J_{\phantom{J},\nu}, \nonumber\\
\Psi^{(2)}_{\mu\nu} & = & \overline{B}_{IK} \delta B^{KL} \overline{B}_{LM} \delta B^{MN} \overline{B}_{NJ} \phi^I_{\phantom{I},\mu} \phi^J_{\phantom{J},\nu}. \nonumber
\end{eqnarray}
This parametrization is applicable up to arbitrary order of the perturbation theory in models with general relativistic solid matter.
When deriving equations of motion for the $n$-th order perturbations all coefficients $C_{abc}$ such that $ a+2b+3c \leq n+1 $ must be specified.

\vskip 1mm
By setting coefficients $C_{abc}$ corresponding to up to second order terms in the equation of state (\ref{eq:eqstate1}) as
\begin{eqnarray}
\label{eq:para1}
& & C_{000} = 1, \quad C_{100} = \frac{w+1}{2}, \\
& & C_{200} = \frac{\lambda+w+1}{8}, \quad C_{010} = \frac{\mu-w-1}{4}, \nonumber
\end{eqnarray}
and taking the Newtonian limit with weak deformation of the solid we retrieve the same equation of state as in theory of elasticity within Newtonian mechanics of continuum \cite{ll}.
Coefficients $\lambda$ and $\mu$ in the equation of state (\ref{eq:eqstate1}) then correspond to Lam\'e parameters $\lambda = \lambda_{\textrm{Lam\'e}}$ and $\mu = \mu_{\textrm{Lam\'e}}$, and $w$ is the pressure to energy density ratio.
In this way we follow \cite{skovran,balek}, but there are also works using different conventions for coefficients in the equation of state for the solid matter \cite{karlovini,bucher}.
In this paper we will study also second order perturbations.
Hence we have to introduce also coefficients $\nu_1$, $\nu_2$ and $\nu_3$, which parametrize the additional coefficients $C_{abc}$ as
\begin{eqnarray}
\label{eq:para2}
C_{110} = \frac{1}{4} \nu_1, \quad C_{001} = \frac{1}{6} \nu_2, \quad C_{300} = \frac{1}{12} \nu_3.
\end{eqnarray}
Other coefficients $C_{abc}$ would be relevant only in higher orders of the perturbation theory, and we may disregard them.

\vskip 1mm
From now on we restrict ourselves to a special case in which the shear stress coefficient $\mu$ vanishes, so that
\begin{eqnarray}
\label{eq:para3}
C_{010} = - \frac{w+1}{4}.
\end{eqnarray}
Such solid matter then does not differ from a perfect fluid when only the background together with the first order perturbations are taken into account.
\vskip 1mm
In the next section we will demand independence of ratio of the time-time component of the mixed stress-energy tensor to its space part trace $T_{0}^{\phantom{0}0}/T_{i}^{\phantom{i}i}$ on the deformation of the solid, i.e. we will make use of validity of the condition (\ref{eq:condition}) in arbitrary coordinates.
This will lead to constraints of coefficient $w$, $\lambda$, $\nu_1$, $\nu_2$ and $\nu_3$.
Of course, one can study more complicated models by relaxing the condition (\ref{eq:condition}), but it is not the aim of this work.

\section{Perturbations in flat universe}\label{sec:3}

We can set any values of coefficients (\ref{eq:para1})-(\ref{eq:para3}) to specify the solid matter, choose any space-time metric, even add other matter components, and study perturbations of such system up to the second order.
However, we will restrict ourselves to a specific case corresponding to homogeneous, isotropic and spatially flat universe, and solid matter satisfying the condition on the mixed tress-energy tensor (\ref{eq:condition}) in arbitrary coordinates.
For the background space-time metric we choose the flat FLRW metric parametrized by cosmic time $t$ and comoving space coordinates $x^i$,
\begin{eqnarray}
ds^2 = - dt^2 + a(t)^2 \delta_{ij} dx^i dx^j,
\end{eqnarray}
with function $a(t)$ denoting the scale factor.
We define the relaxed state of the solid matter by identifying its body coordinates with the comoving space coordinates, $\overline{\phi}^I = \delta^I_i x^i$.
In this way stretching of the solid together with the expanding universe does not count for its deformations in the perturbation theory.
In this section we add perturbations to both space-time metric and solid matter, and derive constraints on coefficients (\ref{eq:para1})-(\ref{eq:para3}) following from the condition (\ref{eq:condition}).

\vskip 1mm
As mentioned above we will restrict ourselves only up to the second order of the perturbation theory.
Due to the linearity in the first order of perturbations, scalar, vector and tensor perturbations are decoupled from each other.
First, we focus on only scalar perturbations even in the second order in which these three kinds of perturbations are coupled.
If we were interested in evolution of scalar perturbations in our universe only up to the second order, such simplification would be reasonable because of two reasons.
The first one is that the first order vector perturbations are decaying in the course of the expansion of the universe \cite{weinberg}, and the second reason is based on observations of the cosmic microwave background radiation which suggest small tensor-to-scalar ratio \cite{planck2}.
Equations for the second order scalar perturbations are sourced by the first order perturbations including vector and tensor ones, but because of their negligibility this source is dominated by scalar perturbations.
Of course, equations for the second order vector and tensor perturbations are sourced in the same manner, and therefore due to presence of the first order scalar perturbations they must be present even if their first order parts are set to zero.
We will include these perturbation later after finding restrictions on coefficients (\ref{eq:para1})-(\ref{eq:para3}) based on analysis of only scalar perturbations.

\vskip 1mm
In general relativistic perturbation theory we are free to make use of small coordinate transformations which do not change the form of the background.
Transformations of tensor $\mathcal{A}$ under such transformations are given by Lie derivatives \cite{stewart}, in particular\cite{abramo}
\begin{eqnarray}
\widetilde{x}^{\mu} = e^{\xi^{\nu} \partial_{\nu}} x^{\mu} \quad \longrightarrow \quad \widetilde{\mathcal{A}} = e^{-\mathcal{L}_{\xi}} \mathcal{A}.
\end{eqnarray}
Setting coordinates in this way is called gauge fixing because of similarity with models with gauge symmetry in other physical contexts.

\vskip 1mm
For scalar sector we use longitudinal gauge in which the perturbed metric is given by two functions $\phi$ and $\psi$,
\begin{eqnarray}
\label{eq:spacetimemetric}
ds^2 = a^2 \left[ -(1+2\phi) d\tau^2 + (1-2\psi) \delta_{ij} dx^i dx^j \right],
\end{eqnarray}
with cosmic time $t$ replaced by the conformal time $\tau=\int a^{-1} dt$.
With this choice the gauge freedom is already exhausted, and the solid matter perturbation has to be parametrized in the most general way.
Therefore we have to add perturbation to the body coordinates $\phi^I$ parametrized by scalar perturbations in the general form,
\begin{eqnarray}
\label{eq:bodycoordinates}
\phi^I = \delta^I_i ( x^i + \pi^i ) = \delta^I_i ( x^i + \sigma_{,i} ),
\end{eqnarray}
but as the last equality reveals, there is only one scalar part of this perturbation, and it is parametrized by function $\sigma$.

\vskip 1mm
When fixing the gauge we can fix the gauge separately for each order of the perturbation theory \cite{matarrese}.
Such order by order gauge fixing simplifies the calculations considerably, however, a dictionary between sets of second order variables with respect to two different ways of gauge fixing contains also terms with first order variables \cite{bruni}.
For convenience, we chose the same gauge for both first and second order perturbations.
Then we can decompose scalar perturbations listed above order by order,
\begin{eqnarray}
\phi = \phi^{(1)} + \phi^{(2)}, \quad \psi = \psi^{(1)} + \psi^{(2)}, \quad \sigma = \sigma^{(1)} + \sigma^{(2)},
\end{eqnarray}\
with the upper indices in parentheses indicating the order of the perturbation theory.

\vskip 1mm
For the stress-energy tensor of the form (\ref{eq:stressenery}) with coefficients in it parametrized as (\ref{eq:para1})-(\ref{eq:para3}), with invariants (\ref{eq:invariants}) defined through body metric (\ref{eq:bodymetric}) corresponding to body coordinates (\ref{eq:bodycoordinates}), and for the space-time metric (\ref{eq:spacetimemetric}), the needed quantities appearing in the condition (\ref{eq:condition}) expressed up to the second order are
\begin{eqnarray}
\label{eq:t00}
\frac{T_0^{\phantom{0}0}}{\overline{\rho}} & = & -1 - 3 (w+1) \psi^{(1)} - (w+1) \triangle \sigma^{(1)} - \\
& & - \frac{3}{2} \left( 5 w + 5 + 3 \lambda  \right) \psi^{(1)2} - 3 (w+1) \psi^{(2)} - \nonumber\\
& & - 3 \left( w + 1 + \lambda \right) \psi^{(1)} \triangle \sigma^{(1)} + \nonumber\\
& & + (w+1) \left( \frac{1}{2} \sigma^{(1)}_{,ij} \sigma^{(1)}_{,ij} - \frac{1}{2} \left( \triangle \sigma^{(1)} \right)^2 - \frac{1}{2} \sigma^{(1)\prime}_{,i} \sigma^{(1)\prime}_{,i} - \triangle \sigma^{(2)} \right) - \nonumber\\
& & - \frac{1}{2} \lambda \left( \triangle \sigma^{(1)} \right)^2, \nonumber
\end{eqnarray}
\begin{eqnarray}
\label{eq:tii}
\frac{T_i^{\phantom{i}i}}{\overline{\rho}} & = & 3w + 9 \lambda \psi^{(1)} + 3 \lambda \triangle \sigma^{(1)} + \\
& & + \frac{3}{2} \left( 15 \lambda + w + 1 + 36 \nu_1 + 8 \nu_2 + 36 \nu_3 \right) \psi^{(1)2} + 9 \lambda \psi^{(2)} + \nonumber\\
& & + \left( w + 1 + 9 \lambda + 36 \nu_1 + 8 \nu_2 + 36 \nu_3 \right) \psi^{(1)} \triangle \sigma^{(1)} + \nonumber\\
& & + (w+1) \left( - \sigma^{(1)}_{,ij} \sigma^{(1)}_{,ij} + \frac{1}{2} \left( \triangle \sigma^{(1)} \right)^2 + \sigma^{(1)\prime}_{,i} \sigma^{(1)\prime}_{,i} \right) + \nonumber\\
& & + \lambda \left( \frac{3}{2} \sigma^{(1)}_{,ij} \sigma^{(1)}_{,ij} + \frac{1}{2} \left( \triangle \sigma^{(1)} \right)^2 - \frac{3}{2} \sigma^{(1)\prime}_{,i} \sigma^{(1)\prime}_{,i} + 3 \triangle \sigma^{(2)} \right) + \nonumber\\
& & + \nu_1 \left( 6 \sigma^{(1)}_{,ij} \sigma^{(1)}_{,ij} + 4 \left( \triangle \sigma^{(1)} \right)^2 \right) + 4 \nu_2 \sigma^{(1)}_{,ij} \sigma^{(1)}_{,ij} + 6 \nu_3 \left( \triangle \sigma^{(1)} \right)^2, \nonumber
\end{eqnarray}
where the prime denotes differentiation with respect to the conformal time.

\vskip 1mm
We can see that satisfying the condition (\ref{eq:condition}) in arbitrary coordinates up to the second order of the perturbation theory is not a trivial matter.
We have to keep track of coefficients at every independent combination of perturbations.
Ratio of a coefficient at a term composed of perturbations in $T_0^{\phantom{0}0}$ to a coefficient at the same perturbation term in $T_i^{\phantom{i}i}$ either must be $-3w$ or both of the coefficients have to be zero.
This must be satisfied for every term in $T_0^{\phantom{0}0}$ and $T_i^{\phantom{i}i}$, which implies a set of constraints on coefficient $w$, $\lambda$, $\nu_1$, $\nu_2$ and $\nu_3$.
It turns out that there is only one solution for this set of constraints.
In this way, by taking into account only the first order parts we find
\begin{eqnarray}
\label{eq:restrict1}
\lambda = w (w+1),
\end{eqnarray}
and by including the second order terms we obtain three additional independent restrictions
\begin{eqnarray}
\label{eq:restrict2}
(3w-1)(w+1) = 0, \quad 3 \nu_1 + 2 \nu_2 = 0, \quad 2 \nu_1 + 3 \nu_3 = 0.
\end{eqnarray}
The first one follows from comparing coefficients at $\sigma^{(1)\prime}_{,i}\sigma^{(1)\prime}_{,i}$, the only term which is zero for static deformations.

\vskip 1mm
In summary, $w=1/3$ and $w=-1$ are the only two values of the pressure to energy density ratio allowing the condition (\ref{eq:condition}) to be satisfied in arbitrary coordinates up to the second order of the perturbation theory with the deformation of the solid dependent on time.
Of course, in the local rest frame the condition (\ref{eq:condition}) holds automatically for any constant value of pressure to energy density ratio, and $\widetilde{w}=w$.
Note also that we have not restricted ourselves to the case with $w=1/3$ or $w=-1$ first, and then studied consequences of the additional condition (\ref{eq:condition}) assumed to be valid in arbitrary coordinates, but we have used only (\ref{eq:condition}).
Therefore, we have shown that validity of (\ref{eq:condition}) implies only $w=1/3$ and $w=-1$ and no other values of the pressure to energy density ratio for the solid matter with zero shear modulus.
The same is true also for a perfect fluid, as shown in appendix \ref{app:a}.

\subsection{Perfect fluid as a special case of solid}

By analysing perturbations up to the second order we have found that only the radiation-like solid with $w=1/3$ and the dark energy-like solid with $w=-1$ satisfy the condition on the mixed stress-energy tensor components (\ref{eq:condition}) in arbitrary coordinates.
The solid with zero shear modulus shares this property with the perfect fluid, as shown in appendix \ref{app:a}.
Let us now approach the fluid in a less usual way, as a special case of the solid.
The perturbative solid matter description is applicable also to an almost homogeneous fluid in an almost homogeneous universe, because elements of the fluid are not displaced too far from their initial comoving positions.

\vskip 1mm
The equation of state for a solid (\ref{eq:eqstate1}) with zero shear modulus and satisfying (\ref{eq:condition}) in arbitrary coordinates which fills a flat FLRW universe written up to the third order in its deformations reduces to
\begin{eqnarray}
\label{eq:eqs00}
\frac{\rho}{\overline{\rho}} & = & 1 + \frac{1}{2} (w+1) \left[ \delta B \right] + \frac{1}{8} \left(w+1\right)^2 \left[ \delta B \right]^2 - \frac{1}{4} (w+1) \left[ \delta B^2 \right] + \\
& & + \nu \left( \frac{1}{4} \left[ \delta B \right] \left[ \delta B^2 \right] - \frac{1}{4} \left[ \delta B^3 \right] - \frac{1}{18} \left[ \delta B \right]^3 \right), \nonumber
\end{eqnarray}
where $\nu_1$ is denoted as $\nu$, $w=1/3$ or $w=-1$, and other coefficients has been written in terms of $w$ and $\nu$.
We can see that for the dark energy-like solid first and second order parts of the equation of state vanish.
It is in principle possible that this special form of equation of state allows no deviation from a perfect fluid.
In such case the description of the continuum through body coordinates and body metric would not be needed at all, and we might end this paper by referring to works studying higher order perturbations in a universe filled with ordinary matter components \cite{cho}.
In order to make sure that this is not the case, and also find out which values of parameters $\nu$ correspond to the perfect fluid, we have to keep using the formalism used so far to compare the equation of state of the perfect fluid with the one for the solid matter.

\vskip 1mm
The energy density of the perfect fluid depends only on volumes of its elements, therefore it is a function of the determinant of the body metric.
It can be written as
\begin{eqnarray}
\frac{\rho_{\textrm{id}}}{\overline{\rho}} = F \left( \textrm{det} B \right) = \sum\limits_{n=0}^{\infty} c_n \left( \textrm{det} \left( \overline{B} B^{-1} \right) - 1 \right)^n,
\end{eqnarray}
where $\overline{B}$ denotes the background body metric, $\overline{B}_{IJ} = a^2 \delta_{IJ}$, and $B^{-1}$ is the inverse body metric containing perturbations.
By expanding this equation of state up to the third order in terms of traces of the body metric perturbations and using the formula for determinant of three by three matrices,
$
\textrm{det} A = \left( \textrm{Tr} A \right)^3 / 6 - \left( \textrm{Tr} A^2 \right) \textrm{Tr} A / 2 + \textrm{Tr} A^3 / 3, 
$
we obtain
\begin{eqnarray}
\frac{\rho_{\textrm{id}}}{\overline{\rho}} & = & c_0 + c_1 \left[ \delta B \right] + \left( \frac{1}{2} c_1 + c_2 \right) \left[ \delta B \right]^2 - \frac{1}{2} c_1 \left[ \delta B^2 \right] + \\
& & + \left( \frac{1}{6} c_1 + c_2 + c_3 \right) \left[ \delta B \right]^3 - \left( \frac{1}{2} c_1 + c_2 \right) \left[ \delta B \right] \left[ \delta B^2 \right] + \frac{1}{3} c_1 \left[ \delta B^3 \right]. \nonumber
\end{eqnarray}
If we compare this equation of state with the equation of state of the solid matter (\ref{eq:eqs00}) we can express coefficients $c_0$-$c_3$ through $w$ and $\nu$ as
\begin{eqnarray}
& & c_0 = 1, \quad c_1 = \frac{1}{2} (w+1), \quad c_2 = \frac{1}{8} (w-1)(w+1), \\
& & c_3 = - \frac{1}{18} \nu - \frac{1}{24} (3w-1)(w+1). \nonumber
\end{eqnarray}
This procedure imposes seven conditions on six parameters $c_0$-$c_3$, $w$ and $\nu$, therefore, we also find restrictions
\begin{eqnarray}
\nu = - \frac{1}{2} (w+1)^2 = - \frac{2}{3} (w+1),
\end{eqnarray}
where the last equality is equivalent to the first condition in (\ref{eq:restrict2}), $(3w-1)(w+1)=0$, which has already been taken into account in (\ref{eq:eqs00}).

\vskip 1mm
In summary, both the radiation-like solid and the dark energy-like solid can be reduced to a perfect fluid.
For the radiation-like case it happens for $\nu = -8/9$, and for the dark energy-like case the reduction occurs when $\nu = 0$.
For other values of coefficient $\nu$ there is difference between the perfect fluid and the solid matter with zero shear modulus, and indeed, in this paper we are studying a non-trivial generalization of the fluid.

\vskip 1mm
Conclusions made so far in this section correspond to only scalar perturbations in the longitudinal gauge.
In order to obtain complete results we also have to examine different choices of the gauge or case without fixing the gauge, and also cases including vector and tensor perturbations.
This is the content of the following subsections.

\subsection{Scalar perturbations without gauge fixing}

Now we relax the longitudinal gauge condition for perturbations.
The FLRW metric with scalar perturbations without any gauge fixing then reads
\begin{eqnarray}
\label{eq:spacetimemetric2}
ds^2 = a^2 \left\{ -\left(1+2\phi\right) d\tau^2 -2B_{,i} d\tau d x^i + \left[ \delta_{ij} \left( 1-2\psi \right) - 2 E_{,ij} \right] d x^i d x^j \right\},
\end{eqnarray}
and considering only scalar perturbations also for the body coordinates we have perturbation $\sigma$ defined in (\ref{eq:bodycoordinates}).
Additional scalar perturbations $B$ and $E$ can by decomposed with respect to order of the perturbation theory in the same way as scalar perturbations used so far, $B = B^{(1)} + B^{(2)}$, $E = E^{(1)} + E^{(2)}$.
We can now repeat calculations leading to results (\ref{eq:t00}) and (\ref{eq:tii}) for the more general form of space-time metric (\ref{eq:spacetimemetric2}), but with constraints on parameters (\ref{eq:restrict1}) and (\ref{eq:restrict2}) already taken into account.

\vskip 1mm
The equation of state constrained by results of analysis of scalar perturbations in the longitudinal gauge up to second order of the perturbation theory (\ref{eq:eqs00}) for the radiation-like case with $w=1/3$ is
\begin{eqnarray}
\label{eq:eqs13}
\frac{\rho}{\overline{\rho}} & = & 1 + \frac{2}{3} \left[ \delta B \right] + \frac{2}{9} \left[ \delta B \right]^2 - \frac{1}{3} \left[ \delta B^2 \right] + \\
& & + \nu \left( \frac{1}{4} \left[ \delta B \right] \left[ \delta B^2 \right] - \frac{1}{4} \left[ \delta B^3 \right] - \frac{1}{18} \left[ \delta B \right]^3 \right), \nonumber
\end{eqnarray}
and the stress-energy tensor corresponding to the space-time metric (\ref{eq:spacetimemetric2}) and the equation of state written above gives
\begin{eqnarray}
\label{eq:417}
\frac{T_0^{\phantom{0}0}}{\overline{\rho}} = - \frac{T_i^{\phantom{i}i}}{\overline{\rho}} & = & -1 - 4 \psi^{(1)} - \frac{4}{3} \triangle E^{(1)} - \frac{4}{3} \triangle \sigma^{(1)} - \\
& & - 4 \psi^{(2)} - \frac{4}{3} \triangle E^{(2)} - 12 \psi^{(1)2} - 8 \psi^{(1)} \triangle E^{(1)} + \nonumber\\
& & + \frac{2}{3} B^{(1)}_{,i} B^{(1)}_{,i} - \frac{4}{3} E^{(1)}_{,ij} E^{(1)}_{,ij} - \frac{8}{9} \left( \triangle E^{(1)} \right)^2 - \nonumber\\
& & - \frac{16}{3} \psi^{(1)} \triangle \sigma^{(1)} - \frac{16}{9} \triangle E^{(1)} \triangle \sigma^{(1)} - \frac{4}{3} \triangle \sigma^{(2)} + \nonumber\\
& & + \frac{2}{3} \sigma^{(1)}_{,ij} \sigma^{(1)}_{,ij} - \frac{8}{9} \left( \triangle \sigma^{(1)} \right)^2 - \frac{2}{3} \sigma^{(1)\prime}_{,i} \sigma^{(1)\prime}_{,i}. \nonumber 
\end{eqnarray}
This is in agreement with the key condition (\ref{eq:condition}).

\vskip 1mm
For the dark energy-like case with $w=-1$ we have
\begin{eqnarray}
\label{eq:eqsm1}
\frac{\rho}{\overline{\rho}} = 1 + \nu \left( \frac{1}{4} \left[ \delta B \right] \left[ \delta B^2 \right] - \frac{1}{4} \left[ \delta B^3 \right] - \frac{1}{18} \left[ \delta B \right]^3 \right),
\end{eqnarray}
and
\begin{eqnarray}
\frac{T_0^{\phantom{0}0}}{\overline{\rho}} = -1, \quad \frac{T_i^{\phantom{i}i}}{\overline{\rho}} = -3,
\end{eqnarray}
so that the condition (\ref{eq:condition}) remains valid in this case as well.

\vskip 1mm
We can conclude that by relaxing the gauge fixing condition which sets the form of the perturbed flat FLRW metric (\ref{eq:spacetimemetric}) we obtain no additional constraints on parameters $w$ and $\nu$ describing the solid obeying the condition (\ref{eq:condition}) assumed to be valid in arbitrary coordinates.

\subsection{Vector and tensor perturbations}\label{sec:43}

If we add vector and tensor perturbations $V_i$ and $\gamma_{ij}$ to the space-time metric using the longitudinal gauge for scalar perturbations and an appropriate gauge for the rest of them, we can write the perturbed metric as
\begin{eqnarray}
\label{eq:mmm}
ds^2 = a^2 \left\{ -\left( 1 + 2 \phi \right) d \tau^2 - 2 V_i d \tau d x^i + \left[ \delta_{ij} \left( 1 - 2 \psi \right) - \gamma_{ij} \right] d x^i d x^j \right\},
\end{eqnarray}
where vector and tensor perturbations satisfy conditions $V_{i,i}=0$, $\gamma_{ii}=0$ and $\gamma_{ij,j}=0$.
We have to add vector perturbation $v^{\perp}_i$ also to the body coordinates,
\begin{eqnarray}
\label{eq:421}
\phi^I = \delta^I_i \left( x^i + v_i \right), \quad v_i = \sigma_{,i} + v^{\perp}_i, \quad v^{\perp}_{i,i} = 0.
\end{eqnarray}
Using order by order gauge fixing we can set the same gauge for the second order perturbations as for the first order ones, and decompose the newly added perturbations in the usual way, $V_i = V^{(1)}_i + V^{(2)}_i$, $\gamma_{ij} = \gamma^{(1)}_{ij} + \gamma^{(2)}_{ij}$, $v^{\perp}_i = v^{\perp(1)}_i + v^{\perp(2)}_i$.

\vskip 1mm
We can assume that $\psi^{(1)} = \phi^{(1)}$, which follows from the traceless part of the Einstein field equations \cite{weinberg,cmbslow}.
Note that this would not be true if we did not set the shear modulus to zero \cite{skovran,balek}.
From the direct calculation of the stress-energy tensor with respect to the metric (\ref{eq:mmm}) for the radiation-like case with $w=1/3$ and the equation of state (\ref{eq:eqs13}) then follows
\begin{eqnarray}
\frac{T_0^{\phantom{0}0}}{\overline{\rho}} = - \frac{T_i^{\phantom{i}i}}{\overline{\rho}} & = & - 1 - 4 \phi^{(1)} - \frac{4}{3} v^{(1)}_{i,i} - 4 \psi^{(2)} - \frac{4}{3} v^{(2)}_{i,i} - \\
& & - 12 \phi^{(1)2} - \frac{16}{3} \phi^{(1)} v^{(1)}_{i,i} + \frac{2}{3} V^{(1)}_{,i} V^{(1)}_{,i} + \nonumber\\
& & + \frac{2}{3} v^{(1)}_{i,j} v^{(1)}_{j,i} - \frac{8}{9} \left( v^{(1)}_{i,i} \right)^2 - \frac{2}{3} v^{(1)\prime}_i v^{(1)\prime}_i - \frac{1}{3} \gamma^{(1)}_{ij} \gamma^{(1)}_{ij}. \nonumber
\end{eqnarray}
Note that this relation contains also scalar perturbation $\sigma$ in the way consistent with the last three terms in (\ref{eq:417}), because according to (\ref{eq:421}) it parametrizes the scalar part of $v_i$.
Similarly, in the dark energy-like case with $w=-1$ and the equation of state (\ref{eq:eqsm1}) we obtain
\begin{eqnarray}
\frac{T_0^{\phantom{0}0}}{\overline{\rho}} = -1, \quad \frac{T_i^{\phantom{i}i}}{\overline{\rho}} = -3.
\end{eqnarray}
In both cases the condition (\ref{eq:condition}) implies the same result as with only the scalar perturbations in the longitudinal gauge. 
Hence the restrictions (\ref{eq:restrict1}) and (\ref{eq:restrict2}) remain valid, and there is no independent restriction which should be added to them.

\section{Scalar perturbations with dark energy-like solid}\label{sec:4}

In this section we focus on the case with $w=-1$, and analyse equations for scalar perturbations up to the second order.
In the case of perfect fluid with $\nu=0$ they should have no solution of the form of perturbations.
We will investigate whether this holds for any value of parameter $\nu$. 

\vskip 1mm
First, we only write down equations and their solution for the unperturbed case, which can be found in standard textbooks.
The nontrivial Einstein equations in the zeroth order of the perturbation theory are
\begin{eqnarray}
& & G^{(0)}_{00} = 3 \mathcal{H}^2 = 8\pi\kappa T^{(0)}_{00} = 8\pi\kappa a^2 \overline{\rho}, \\
& & G^{(0)}_{ij} = - \left( \mathcal{H}^2 + 2 \mathcal{H}^{\prime} \right) \delta_{ij} = 8\pi\kappa T^{(0)}_{ij} = - 8\pi\kappa a^2 \overline{\rho} \delta_{ij},
\end{eqnarray}
where $\mathcal{H}= a^{\prime} / a$, and the only nontrivial part of the energy-momentum conservation law is
\begin{eqnarray}
T^{(0)0\mu}_{\phantom{(0)0\mu};\mu} = \frac{\overline{\rho}^{\prime}}{a^2} = 0.
\end{eqnarray}
These equations are solved for
\begin{eqnarray}
\label{eq:background1}
a \propto \frac{1}{\tau}, \quad \overline{\rho} = \textrm{const.},
\end{eqnarray}
hence $\mathcal{H} = - 1 / \tau$.
We will use this result as the background for higher orders of the perturbation theory.

\subsection{Linear perturbations}

Not only the background equations, but also equations for the linearised cosmological perturbations can be found in the standard textbooks.
Here we are including the first order equations for two reasons.
The first one is that they are a step towards analysis of the second order perturbations which will be studied in the next subsection.
The second reason is that textbooks tend to omit them in the case with dark energy-like pressure to energy density ratio $w=-1$, which is justified by lack of any nontrivial solution, as we will see bellow.
The needed equations in the first order of the perturbation theory are
\begin{eqnarray}
\label{eq:00}
& & G^{(1)}_{00} = 2 \left( \triangle \phi^{(1)} - 3 \mathcal{H} \phi^{(1)\prime} \right) = 8\pi\kappa T^{(1)}_{00} = 6 \mathcal{H}^2 \phi^{(1)}, \\
\label{eq:0i}
& & G^{(1)}_{0i} = 2 \left( \mathcal{H} \phi^{(1)}_{,i} + \phi^{(1)\prime}_{,i} \right) = 8\pi\kappa T^{(1)}_{0i} = 0, \\
\label{eq:ij}
& & G^{(1)}_{ij} = 2 \left( 6 \mathcal{H}^2 \phi^{(1)} + 3 \mathcal{H} \phi^{(1)\prime} + \phi^{(1)\prime\prime} \right) \delta_{ij} = 8\pi\kappa T^{(1)}_{ij} = 6 \mathcal{H}^2 \phi^{(1)} \delta_{ij},
\end{eqnarray}
with the energy-momentum conservation law $T^{(1)\mu\nu}_{\phantom{(1)\mu\nu};\nu}=0$ implying only $0=0$ type of equations.
These equations has been simplified with the use of the background solution (\ref{eq:background1}), $\mathcal{H}^{\prime}=\mathcal{H}^2$, and the fact that $\psi^{(1)} = \phi^{(1)}$ which follows from the traceless spatial part of Einstein field equations, $(\phi^{(1)}-\psi^{(1)})_{,ij} \stackrel{i \neq j}{=}0$ together with the assumption that $\phi^{(1)}$ and $\psi^{(1)}$ are of the form of perturbations.
The latter simplification has already been used in subsection \ref{sec:43}.

\vskip 1mm
The Fourier mode of perturbation $\phi^{(1)}$ corresponding to the solution of the $00$-part of the Einstein field equations (\ref{eq:00}) is
\begin{eqnarray}
\label{eq:f1}
\phi^{(1)}_k = \alpha_1 \tau e^{\frac{1}{6}k^2\tau^2},
\end{eqnarray}
for the $0i$-part (\ref{eq:0i}) it is
\begin{eqnarray}
\label{eq:f2}
\phi^{(1)}_k = \alpha_2 \tau + \tau \int\limits_{\tau_0}^{\tau} \frac{ d \tau^{\prime}}{\tau^{\prime}} f\left(\tau^{\prime}\right),
\end{eqnarray}
and from the $ij$-part (\ref{eq:ij}) follows
\begin{eqnarray}
\label{eq:f3}
\phi^{(1)}_k = \alpha_3\tau+\alpha_4\tau^3,
\end{eqnarray}
where $\alpha_1$-$\alpha_4$ and $\tau_0$ are constants and $f$ is an arbitrary function.
Consistency of three forms of the solution for the scalar perturbation $\phi^{(1)}$ (\ref{eq:f1})-(\ref{eq:f3}) can be satisfied only when $\alpha_1=\alpha_2=\alpha_3$, $\alpha_4=0$, $f=0$ and $\tau_0=0$ and only for mode with zero wavenumber, so that $\phi^{(1)}=\alpha_1\tau$.

\vskip 1mm
The found $\phi^{(1)}$ is not of a form of perturbation and can be absorbed into the background.
The metric corresponding to this ''new'' background can be written as
\begin{eqnarray}
ds^{2(0)}_{\textrm{new}} = \frac{1}{\tau^2} \left[ -\left(1-\varepsilon\tau\right)d\tau^2+\left(1+\varepsilon\tau\right)\delta_{ij}dx^idx^j \right],
\end{eqnarray}
where the scale factor is chosen to be $a=1$ at the conformal time $\tau=1$, and parameter $\varepsilon=-2\alpha_1$ must be small.
It turns out that this ''new'' background metric does not differ from the original background metric, because by mere redefinition of the conformal time $\widetilde{\tau}=\tau-\varepsilon\tau^2/2+\mathcal{O}(\varepsilon^2)$ we find
\begin{eqnarray}
ds^{2(0)}_{\textrm{new}} = \frac{1}{\widetilde{\tau}^2}\left( -d\widetilde{\tau}^2+\delta_{ij}dx^idx^j \right)+\mathcal{O}(\varepsilon^2).
\end{eqnarray}
Hence in the universe with dark energy the first order perturbations must be omitted.

\vskip 1mm
We can also see that perturbation $\sigma^{(1)}$ corresponding to body element displacement does not appear in equations for the first order perturbations due to its absence in the stress-energy tensor, and thus, $\sigma^{(1)}$ remains arbitrary.
In other words, in the dark energy-like case the first order part of perturbation $\sigma$ is a redundant quantity.

\subsection{Second order perturbations}

Do the second order perturbations vanish not only, as expected in the case with the dark energy as a perfect fluid, for $\nu=0$, but also for dark energy-like solid with zero shear modulus and with $\nu\neq 0$?
A naive answer would be yes, because the condition (\ref{eq:condition}) assumed to be valid in arbitrary coordinates holds for any value of parameter $\nu$, and in the dark energy-like case it is related to stress-energy tensor proportional to the space-time metric.
However, $T_{\mu\nu}\propto g_{\mu\nu}$ implies both validity of condition (\ref{eq:condition}) in arbitrary coordinates and absence of matter perturbation $\sigma$ in equations in any perturbative order, the opposite is not true, because the condition (\ref{eq:condition}) does not necessarily imply $T_{\mu\nu}\propto g_{\mu\nu}$.
With nonzero coefficient $\nu$ we depart from the perfect fluid case, and there appears traceless space part of the stress energy tensor, which does not violate the condition (\ref{eq:condition}), but it may affect the evolution of perturbations.

\vskip 1mm
Let us examine equations for the second order perturbations.
The corresponding Einstein field equations and the energy-momentum conservation law read
\begin{eqnarray}
G^{(2)}_{00} & = & 2 \triangle \psi^{(2)} - 6 \mathcal{H} \psi^{(2)\prime} + \\
& & + 3\phi^{(1)}_{,i}\phi^{(1)}_{,i} + 12\phi^{(1)}\triangle\phi^{(1)} - 12\mathcal{H}\phi^{(1)}\phi^{(1)\prime} + \phi^{(1)\prime 2} = \nonumber\\
& = & 8\pi\kappa T^{(2)}_{00} = 6\mathcal{H}^2\phi^{(2)}, \nonumber\\
G^{(2)}_{0i} & = & 2\mathcal{H}\phi^{(2)}_{,i} + 2\psi^{(2)\prime}_{,i} - \\
& & - 4\mathcal{H}\phi^{(1)}\phi^{(1)}_{,i} + 2\phi^{(1)}_{,i}\phi^{(1)\prime} + 4\phi^{(1)}\phi^{(1)\prime}_{,i} = \nonumber\\
& = & 8\pi\kappa T^{(2)}_{0i} = 0, \nonumber\\
G^{(2)}_{ii} & = & 6\psi^{(2)\prime\prime} + 6\mathcal{H}\left(\phi^{(2)}+2\psi^{(2)}\right)^{\prime} + \\
& & +18\mathcal{H}^2\left(\phi^{(2)}+\psi^{(2)}\right) + 2\triangle\left(\phi^{(2)}-\psi^{(2)}\right) - \nonumber\\
& & - 72\mathcal{H}^2\phi^{(1)2} - 7\phi^{(1)}_{,i}\phi^{(1)}_{,i} - 8\phi^{(1)}\triangle\phi^{(1)} - \nonumber\\
& & - 60\mathcal{H}\phi^{(1)}\phi^{(1)\prime} - 3\phi^{(1)\prime 2} - 12\phi^{(1)}\phi^{(1)\prime\prime} = \nonumber\\
& = & 8\pi\kappa T^{(2)}_{ii} = 18\mathcal{H}^2\psi^{(2)}, \nonumber\\
G^{(2)}_{ij} & \stackrel{i\neq j}{=} & - \left(\phi^{(2)}-\psi^{(2)}\right)_{,ij} + 2\phi^{(1)}_{,i}\phi^{(1)}_{,j} + 4\phi^{(1)}\phi^{(1)}_{,ij} \stackrel{i\neq j}{=} \\
& \stackrel{i\neq j}{=} & 8\pi\kappa T^{(2)}_{ij} \stackrel{i\neq j}{=} 3\mathcal{H}^2\nu\left(-6\sigma^{(1)}_{,ik}\sigma^{(1)}_{,jk}+4\sigma^{(1)}_{,ij}\triangle\sigma^{(1)}\right), \nonumber\\
\label{eq:517}
T^{(2)i\mu}_{\phantom{(2)i\mu};\mu} & = & \frac{\nu\overline{\rho}}{a^2}\left( \frac{4}{3}\triangle\sigma^{(1)}_{,i}\triangle\sigma^{(1)} - 2\sigma^{(1)}_{,ij}\triangle\sigma^{(1)}_{,j} - 2 \sigma^{(1)}_{,ijk}\sigma^{(1)}_{,jk} \right) = 0,
\end{eqnarray}
where these relations are simplified with the use of the background solution together with relation $\psi^{(1)}=\phi^{(1)}$ in the same way as the first order equations, $ij$-part of the Einstein field equations is split into its trace and traceless part, and equation $T^{(2)0\mu}_{\phantom{(2)0\mu};\mu}=0$ is automatically satisfied.
Since $\phi^{(1)}=0$, the second order Einstein field equations reduce to
\begin{eqnarray}
& & \triangle\psi^{(2)} - 3\mathcal{H}\psi^{(2)\prime} = 3\mathcal{H}^2\phi^{(2)}, \\
& & \mathcal{H}\phi^{(2)} + \psi^{(2)\prime} = f^{(2)}, \\
& & \psi^{(2)\prime\prime} + \mathcal{H}\left(\phi^{(2)}+2\psi^{(2)}\right)^{\prime} + 3\mathcal{H}^2\phi^{(2)} + \frac{1}{3}\triangle\left(\phi^{(2)}-\psi^{(2)}\right) = 0, \\
& & \left(\phi^{(2)}-\psi^{(2)}\right)_{ij} \stackrel{i\neq j}{=} 3 \nu \mathcal{H}^2\left( 6\sigma^{(1)}_{,ik}\sigma^{(1)}_{,jk} - 4\sigma^{(1)}_{,ij}\triangle\sigma^{(1)} \right),
\end{eqnarray}
with $f^{(2)}$ denoting any function of only time.
In other words, equations for the second order scalar perturbations are not sourced by their first order parts, because they vanish, and here we are neglecting first order vector and tensor perturbations.

\vskip 1mm
The second order momentum conservation law (\ref{eq:517}) imposes a restriction on the first order perturbation $\sigma^{(1)}$ which, assuming that it is of the form of a perturbation, is satisfied only if this perturbation vanishes.
As a consequence of vanishing $\sigma^{(1)}$ the system of equations for the second order perturbations reduces to the same system of equations as for the first order perturbations.
This means that the metric perturbation $\phi$ vanishes not only in the first order but also in the second order of the perturbation theory.

\vskip 1mm
The fourth order conservation law $T^{(4)i\mu}_{\phantom{(4)i\mu};\mu}=0$ contains the same combinations of $\sigma^{(2)}$ as combinations of $\sigma^{(1)}$ appearing in the second order equation $T^{(2)i\mu}_{\phantom{(4)i\mu};\mu}=0$, and in principle only such combinations of other perturbations that follow the structure $\phi^{(1)}\phi^{(3)}$, $\phi^{(2)2}$, $\sigma^{(1)}\phi^{(3)}$, $\phi^{(1)}\sigma^{(3)}$ and $\sigma^{(1)}\sigma^{(3)}$, which are zero.
This implies the same restriction on $\sigma^{(2)}$ as on $\sigma^{(1)}$.
In the case of the perfect fluid no such restriction emerges, so that perturbation $\sigma^{(1)}$ remains unspecified.
This means that for the dark energy as a perfect fluid the body element displacement $\sigma_{,i}$ does not affect the evolution of the metric perturbation $\phi$, which turns out to be zero as well, and entire solid matter formalism can be simply replaced by the standard perfect fluid approach.
On the other hand, in the case with dark energy as a solid with zero shear stress and $\nu \neq 0$ this body element approach is needed, and it helps to reveal that perturbation $\sigma$ vanishes.

\vskip 1mm
Therefore, regardless of choice of the parameter $\nu$ there are no scalar perturbations in the dark energy-like solid up to the second order of the perturbation theory.
Moreover, due to the absence of perturbation $\sigma$ in the stress-energy tensor the dark energy perturbations can be omitted also in more realistic multicomponent cosmological models, as long as there is no direct interaction of the dark energy sector with other matter components of the universe.

\vskip 1mm
The analysis presented in this section shows that by considering only scalar perturbations, presence of only dark energy in the expanding universe does not allow for their formation up to the second perturbative order in both standard dark energy model as well as the extended model studied here. The same is true also for vector perturbations, however, tensor perturbations are formed already in the linear order, where is no difference between the standard dark energy and our extension of it. In both models the Fourier modes of tensor perturbations in the first order of the perturbation theory are given by Bessel functions of the order $3/2$, in particular $(-k\tau)J_{3/2}(-k\tau)$ or $(-k\tau)Y_{3/2}(-k\tau)$, which have nondecaying parts. Note that in the context of slow-roll inflation, in which condition for pressure to energy density ratio $w=-1$ must be violated in order to avoid eternal inflation and allow formation of scalar perturbations, these modes are slightly modified by slow-roll parameters.

\vskip 1mm
In spite of lacking a difference in results for up to second order scalar perturbations between our dark energy model and the standard dark energy model, such difference can be pursued by investigating terms quadratic in first order tensor perturbations in equations for second order perturbations, where they play the role of source terms. In this paper we omitted this possibility motivated by limit on tensor to scalar ratio suggested by observations, so that in such context our calculations represent only the leading order approximation. Another exciting possibility, already hinted above, is to extend our dark energy model even more by considering nonminimal coupling to other kinds of matter or to gravity. This would, of course, complicate the analysis considerably.

\section{Scalar perturbations with radiation-like solid}\label{sec:5}

We now examine scalar perturbations in the case with $w=1/3$.
Up to the first order of the perturbation theory we only reproduce results known from textbooks \cite{weinberg}, but we also reveal the way in which parameter $\nu$ enters formulas for second order perturbations.

\vskip 1mm
The only nontrivial background equations are
\begin{eqnarray}
& & G^{(0)}_{00} = 3\mathcal{H}^2 = 8\pi\kappa T^{(0)}_{00} = 8\pi\kappa a^2\overline{\rho}, \\
& & G^{(0)}_{ij} = - \left( \mathcal{H}^2 + 2\mathcal{H}^{\prime} \right) \delta_{ij} = 8\pi\kappa T^{(0)}_{ij} = \frac{8}{3}\pi\kappa a^2\overline{\rho} \delta_{ij}, \\
& & T^{(0)0\mu}_{\phantom{(0)0\mu};\mu} = \frac{\overline{\rho}^{\prime} + 4\mathcal{H}\overline{\rho}}{a^2},
\end{eqnarray}
with
\begin{eqnarray}
\label{eq:background2}
a \propto \tau, \quad \overline{\rho} \propto \frac{1}{\tau^4},
\end{eqnarray}
being solution of these equations, so that $\mathcal{H} = 1/\tau$.

\subsection{Linear perturbations}

Using the background solution (\ref{eq:background2}) and the fact that $\psi^{(1)}=\phi^{(1)}$ which follows from the fist order traceless spatial part of Einstein field equations and has already been used twice in this paper, we can write all nontrivial first order equations as
\begin{eqnarray}
& & G^{(1)}_{00} = 2 \left( \triangle \phi^{(1)} - 3 \mathcal{H} \phi^{(1)\prime} \right) = 8\pi\kappa T^{(1)}_{00} = 2 \mathcal{H}^2 \left( 2\triangle\sigma^{(1)} + 9 \phi^{(1)} \right), \\
& & G^{(1)}_{0i} = 2 \left( \mathcal{H} \phi^{(1)}_{,i} + \phi^{(1)\prime}_{,i} \right) = 8\pi\kappa T^{(1)}_{0i} = 4\mathcal{H}^2\sigma^{(1)\prime}_{,i}, \\
& & G^{(1)}_{ij} = 2 \left( -2 \mathcal{H}^2 \phi^{(1)} + 3 \mathcal{H} \phi^{(1)\prime} + \phi^{(1)\prime\prime} \right) \delta_{ij} = \\
& & \phantom{G^{(1)}_{ij}} = 8\pi\kappa T^{(1)}_{ij} = \mathcal{H}^2 \left( \frac{4}{3} \triangle\sigma^{(1)} + 2 \phi^{(1)} \right) \delta_{ij}. \nonumber
\end{eqnarray}
By combining the $00$-part of the Einstein field equations with the $ij$-part we can exclude the perturbation $\sigma^{(1)}$ and write down the equation for the metric perturbation
\begin{eqnarray}
\phi^{(1)\prime\prime} + 4\mathcal{H}\phi^{(1)\prime} - \frac{1}{3}\triangle\phi^{(1)} = 0,
\end{eqnarray}
and then supplement the $0i$-part with its solution to find the solution for perturbation $\sigma^{(1)}$ as well.
In this way we find Fourier modes
\begin{eqnarray}
\phi^{(1)}_k & = & \frac{c_1}{\tau^2} \left( \frac{\cos\zeta}{\zeta} + \sin\zeta \right) + \frac{c_2}{\tau^2} \left( \frac{\sin\zeta}{\zeta} - \cos\zeta \right), \\
\sigma^{(1)}_k & = & c_1 \left( \frac{\cos\zeta}{\zeta} + \frac{\sin\zeta}{2} \right) + c_2 \left( \frac{\sin\zeta}{\zeta} - \frac{\cos\zeta}{2} \right), \quad \zeta\equiv\frac{k\tau}{\sqrt{3}},
\end{eqnarray}
where we see that the corresponding sound speed is $1/\sqrt{3}$ as expected for radiation.
Because of the Bianchi identity this solution also satisfies the energy-momentum conservation law,
\begin{eqnarray}
T^{(1)i\mu}_{\phantom{(1)i\mu};\mu} = \frac{4\overline{\rho}}{3a^2} \left( 2\phi^{(1)}_{,i} + \frac{1}{3}\triangle\sigma^{(1)}_{,i} - \sigma^{(1)\prime\prime}_{,i} \right) = 0.
\end{eqnarray}

\subsection{Second order perturbations}

List of the needed equations for the second order perturbations is
\begin{eqnarray}
G^{(2)}_{00} & = & 2 \left( \triangle\psi^{(2)} - 3\mathcal{H}\psi^{(2)\prime} \right) + \\
& & + 3\phi^{(1)}_{,i}\phi^{(1)}_{,i} + 12\phi^{(1)}\triangle\phi^{(1)} - 12\mathcal{H}\phi^{(1)}\phi^{(1)\prime} + 3\phi^{(1)\prime 2} = \nonumber\\
& = & 8\pi\kappa T^{(2)}_{00} = 2 \mathcal{H}^2 \bigg( 2\triangle\sigma^{(2)} + 3\left(\phi^{(2)}+2\psi^{(2)}\right) + \nonumber\\
& & + 30\phi^{(1)2} + 12\phi^{(1)}\triangle\sigma^{(1)} + \frac{4}{3}\left(\triangle\sigma^{(1)}\right)^2 - \sigma^{(1)}_{,ij}\sigma^{(1)}_{,ij} + \sigma^{(1)\prime}_{,i}\sigma^{(1)\prime}_{,i} \bigg), \nonumber\\
G^{(2)}_{0i} & = & 2\mathcal{H}\phi^{(2)}_{,i} + 2\psi^{(2)\prime}_{,i} - \\
& & - 4\mathcal{H}\phi^{(1)}\phi^{(1)}_{,i} + 2\phi^{(1)}_{,i}\phi^{(1)\prime} + 4\phi^{(1)}\phi^{(1)\prime}_{,i} = \nonumber\\
& = & 8\pi\kappa T^{(2)}_{0i} = \mathcal{H}^2 \bigg( 4\sigma^{(2)\prime}_{,i} + \nonumber\\
& & + 8\phi^{(1)}\sigma^{(1)\prime}_{,i} + \frac{16}{3} \sigma^{(1)\prime}_{,i}\triangle\sigma^{(1)} - 4\sigma^{(1)}_{,ij}\sigma^{(1)\prime}_{,ij} \bigg), \nonumber\\
G^{(2)}_{ii} & = & 6\psi^{(2)\prime\prime} + 6\mathcal{H}\left(\phi^{(2)}+2\psi^{(2)}\right)^{\prime} - \\
& & - 6\mathcal{H}^2\left(\phi^{(2)}+\psi^{(2)}\right) + 2\triangle\left(\phi^{(1)}-\psi^{(1)}\right) + \nonumber\\
& & + 24\mathcal{H}^2\phi^{(1)2} - 7\phi^{(1)}_{,i}\phi^{(1)}_{,i} - 8\phi^{(1)}\triangle\phi^{(1)} - \nonumber\\
& & - 60\mathcal{H} \phi^{(1)}\phi^{(1)\prime} - 3\phi^{(1)\prime 2} - 12\phi^{(1)}\phi^{(1)\prime\prime} = \nonumber\\
& = & 8\pi\kappa T^{(2)}_{ii} = \mathcal{H}^2 \bigg( 4\triangle\sigma^{(2)} + 6\psi^{(2)} + \nonumber\\
& & + 12\phi^{(1)2} + 8\phi^{(1)}\triangle\sigma^{(1)} - 2\sigma^{(1)}_{,ij}\sigma^{(1)}_{,ij} + \frac{8}{3}\left(\triangle\sigma^{(1)}\right)^2 + 2\sigma^{(1)\prime}_{,i}\sigma^{(1)\prime}_{,i} \bigg), \nonumber\\
G^{(2)}_{ij} & \stackrel{i\neq j}{=} & - \left(\phi^{(2)}-\psi^{(2)}\right)_{,ij} + 2\phi^{(1)}_{,i}\phi^{(1)}_{,j} + 4\phi^{(1)}\phi^{(1)}_{,ij} \stackrel{i\neq j}{=} \\
& \stackrel{i\neq j}{=} & 8\pi\kappa T^{(2)}_{ij} \stackrel{i\neq j}{=} \mathcal{H}^2 \bigg[ -16\left(1+\frac{9}{8}\nu\right)\sigma^{(1)}_{,ik}\sigma^{(1)}_{,jk} + \nonumber\\
& & + \frac{32}{3}\left(1+\frac{9}{8}\nu\right)\sigma^{(1)}_{,ij}\triangle\sigma^{(1)} + 4\sigma^{(1)\prime}_{,i}\sigma^{(1)\prime}_{,j} \bigg], \nonumber\\
T^{(2)0\mu}_{\phantom{(2)0\mu};\mu} & = & \frac{4\overline{\rho}}{3a^2} \left( -2\phi^{(1)}_{,i}\sigma^{(1)\prime}_{,i} - \frac{1}{3}\sigma^{(1)\prime}_{,i}\triangle\sigma^{(1)}_{,i} + \sigma^{(1)\prime}_{,i}\sigma^{(1)\prime\prime}_{,i} \right), \\
T^{(2)i\mu}_{\phantom{(2)i\mu};\mu} & = & \frac{4\overline{\rho}}{3a^2} \bigg[ \left(\phi^{(2)}+\psi^{(2)}\right)_{,i} + \frac{1}{3}\triangle\sigma^{(2)}_{,i} - \sigma^{(2)\prime\prime}_{,i} + 12\phi^{(1)}\phi^{(1)}_{,i} + \\
& & + \frac{8}{3}\phi^{(1)}_{,i}\triangle\sigma^{(1)} + 2\phi^{(1)}\triangle\sigma^{(1)}_{,i} + 2\phi^{(1)\prime}\sigma^{(1)\prime}_{,i} -2\phi^{(1)}\sigma^{(1)\prime\prime}_{,i} + \nonumber\\
& & + \frac{5}{3} \sigma^{(1)\prime}_{,ij}\sigma^{(1)\prime}_{,j} - \frac{1}{3}\sigma^{(1)\prime}_{,i}\triangle\sigma^{(1)\prime} + \sigma^{(1)}_{,ij}\sigma^{(1)\prime\prime}_{,j} - \frac{4}{3}\sigma^{(1)\prime\prime}_{,i}\triangle\sigma^{(1)} + \nonumber\\
& & + \left(\frac{4}{3}+\nu\right)\triangle\sigma^{(1)}_{,i}\triangle\sigma^{(1)} -\left(\frac{5}{3}+\frac{3}{2}\nu\right)\sigma^{(1)}_{,ijk}\sigma^{(1)}_{,jk} - \nonumber\\
& & -\left(\frac{4}{3}+\frac{3}{2}\nu\right)\sigma^{(1)}_{,ij}\triangle\sigma^{(1)}_{,j} \bigg]. \nonumber
\end{eqnarray}
In order to find the solution we can rewrite the system of equations as
\begin{eqnarray}
\label{eq:first}
T^{(2)i\mu}_{\phantom{(2)i\mu};\mu,i} = 0 & \implies & \phi^{(2)}+\psi^{(2)}+\frac{1}{3}\triangle\sigma^{(2)}-\sigma^{(2)\prime\prime} = \mathcal{Q}_1, \\
\label{eq:second}
G^{(2)}_{00} = 8\pi\kappa T^{(2)}_{00} & \implies & \phi^{(2)}+2\psi^{(2)}+\frac{1}{\mathcal{H}}\psi^{(2)\prime}- \\
& & - \frac{1}{3\mathcal{H}^2}\triangle\psi^{(2)}+\frac{2}{3}\triangle\sigma^{(2)} = \mathcal{Q}_2, \nonumber\\
\label{eq:third}
G^{(2)\phantom{0}0}_{\phantom{(2)}0}+G^{(2)\phantom{i}i}_{\phantom{(2)}i} = 0 & \implies & \psi^{(2)\prime\prime} + \mathcal{H}\left(\phi^{(2)\prime}+3\psi^{(2)\prime}\right) + \\
\label{eq:fourth}
& & + \frac{1}{3}\triangle\left(\phi^{(2)}-2\psi^{(2)}\right) = \mathcal{Q}_3, \nonumber\\
G^{(2)}_{0i,i} = 8\pi\kappa T^{(2)}_{0i,i} & \implies & \frac{1}{\mathcal{H}^2}\psi^{(2)\prime} + \frac{1}{\mathcal{H}}\phi^{(2)} - 2\sigma^{(2)\prime} = \mathcal{Q}_4,
\end{eqnarray}
with $G^{(2)\phantom{0}0}_{\phantom{(2)}0}+G^{(2)\phantom{i}i}_{\phantom{(2)}i}=0$ being valid only when condition (\ref{eq:condition}) is satisfied with radiation-like pressure to energy density ratio $w=1/3$, and $\mathcal{Q}_1$-$\mathcal{Q}_4$ denoting the source terms of these equations,
\begin{eqnarray}
\mathcal{Q}_1 & = & \triangle^{-1}\bigg[ -12\phi^{(1)}_{,i}\phi^{(1)}_{,i} - 12\phi^{(1)}\triangle\phi^{(1)} - \\
& & - \frac{8}{3} \triangle\phi^{(1)}\triangle\sigma^{(1)} - \frac{14}{3}\phi^{(1)}_{,i}\triangle\sigma^{(1)}_{,i} - 2\phi^{(1)}\triangle^2\sigma^{(1)} - \nonumber\\
& & - 2\phi^{(1)\prime}_{,i}\sigma^{(1)\prime}_{,i} - 2\phi^{(1)\prime}\triangle\sigma^{(1)\prime} + 2\phi^{(1)}_{,i}\sigma^{(1)\prime\prime}_{,i} + 2\phi^{(1)}\triangle\sigma^{(1)\prime\prime} - \nonumber\\
& & - \frac{4}{3} \sigma^{(1)\prime}_{,i}\triangle\sigma^{(1)\prime}_{,i} - \frac{5}{3}\sigma^{(1)\prime}_{,ij}\sigma^{(1)\prime}_{,ij} + \frac{1}{3}\left(\triangle\sigma^{(1)\prime}\right)^2 + \nonumber\\
& & + \frac{1}{3}\sigma^{(1)\prime\prime}_{,i}\triangle\sigma^{(1)}_{,i} - \sigma^{(1)\prime\prime}_{,ij}\sigma^{(1)}_{,ij} + \frac{4}{3}\triangle\sigma^{(1)\prime\prime}\triangle\sigma^{(1)} - \nonumber\\
& & - \left(\frac{4}{3}+\nu\right)\triangle\sigma^{(1)}\triangle^2\sigma^{(1)} + \frac{1}{2}\nu\triangle\sigma^{(1)}_{,i}\triangle\sigma^{(1)}_{,i} + \nonumber\\
& & + \left(3+3\nu\right)\sigma^{(1)}_{,ij}\triangle\sigma^{(1)}_{,ij} + \left(\frac{5}{3}+\frac{3}{2}\nu\right)\sigma^{(1)}_{,ijk}\sigma^{(1)}_{,ijk} \bigg], \nonumber\\
\mathcal{Q}_2 & = & - 10\phi^{(1)2} - \frac{2}{\mathcal{H}}\phi^{(1)}\phi^{(1)\prime} + \frac{1}{2\mathcal{H}^2}\phi^{(1)\prime 2} + \frac{1}{2\mathcal{H}^2}\phi^{(1)}_{,i}\phi^{(1)}_{,i} + \frac{2}{\mathcal{H}^2}\phi^{(1)}\triangle\phi^{(1)} - \\
& & - 4\phi^{(1)}\triangle\sigma^{(1)} - \frac{4}{9}\left(\triangle\sigma^{(1)}\right)^2 + \frac{1}{3}\sigma^{(1)}_{,ij}\sigma^{(1)}_{,ij} - \frac{1}{3}\sigma^{(1)\prime}_{,i}\sigma^{(1)\prime}_{,i}, \nonumber\\
\mathcal{Q}_3 & = & \phi^{(1)\prime 2} + \frac{5}{3}\phi^{(1)}_{,i}\phi^{(1)}_{,i} + \frac{8}{3}\phi^{(1)}\triangle\phi^{(1)} + 4\mathcal{H}\phi^{(1)}\phi^{(1)\prime}, \\
\mathcal{Q}_4 & = & \triangle^{-1} \bigg[ \frac{2}{\mathcal{H}}\phi^{(1)}_{,i}\phi^{(1)}_{,i} + \frac{2}{\mathcal{H}}\phi^{(1)}\triangle\phi^{(1)} - \frac{1}{\mathcal{H}^2}\triangle\phi^{(1)}\phi^{(1)\prime} - \\
& & - \frac{3}{\mathcal{H}^2}\phi^{(1)}_{,i}\phi^{(1)\prime}_{,i} - \frac{2}{\mathcal{H}^2}\phi^{(1)}\triangle\phi^{(1)\prime} + 4\phi^{(1)}_{,i}\sigma^{(1)\prime}_{,i} + 4\phi^{(1)}\triangle\sigma^{(1)\prime} + \nonumber\\
& & + \frac{8}{3}\triangle\sigma^{(1)\prime}\triangle\sigma^{(1)} + \frac{2}{3}\sigma^{(1)\prime}_{,i}\triangle\sigma^{(1)}_{,i} - 2\sigma^{(1)\prime}_{,ij}\sigma^{(1)}_{,ij} \bigg]. \nonumber
\end{eqnarray}
Here we have assumed that the source terms are dominated by scalar perturbations.
By combining equation (\ref{eq:first}) with the conformal time derivative of (\ref{eq:fourth}) we find
\begin{eqnarray}
\label{eq:sigma}
\frac{1}{3} \triangle \sigma^{(2)} = \mathcal{Q}_1 - \frac{1}{2} \mathcal{Q}_4^{\prime} - \frac{1}{2} \phi^{(2)} - \psi^{(2)} + \frac{1}{\mathcal{H}} \left( \frac{1}{2} \phi^{(2)\prime} + \psi^{(2)\prime} \right) + \frac{1}{2\mathcal{H}^2} \psi^{(2)\prime\prime}.
\end{eqnarray}
By combining this equation with (\ref{eq:second}) we obtain
\begin{eqnarray}
- \psi^{(2)\prime\prime} - \mathcal{H} \left( \phi^{(2)\prime} + 3 \psi^{(2)\prime} \right) + \frac{1}{3} \triangle \psi^{(1)} = \mathcal{H}^2 \left( 2 \mathcal{Q}_1 - \mathcal{Q}_2 - \mathcal{Q}_4^{\prime} \right),
\end{eqnarray}
and after adding this equation to (\ref{eq:third}) we find the relation
\begin{eqnarray}
\label{eq:fipsi}
\phi^{(2)} - \psi^{(2)} = 3 \triangle^{-1} \left[ \mathcal{Q}_3 + \mathcal{H}^2 \left( 2 \mathcal{Q}_1 - \mathcal{Q}_2 - \mathcal{Q}_4^{\prime} \right) \right].
\end{eqnarray}
By inserting $\phi^{(2)}$ from this equation back into (\ref{eq:third}) we have
\begin{eqnarray}
\psi^{(2)\prime\prime} + 4 \mathcal{H} \psi^{(2)\prime} - \frac{1}{3} \triangle \psi^{(2)} = \mathcal{Q}_0,
\end{eqnarray}
where
\begin{eqnarray}
\mathcal{Q}_0 = - \mathcal{H}^2 \left( 2 \mathcal{Q}_1 - \mathcal{Q}_2 - \mathcal{Q}_4^{\prime} \right) - 3 \mathcal{H} \triangle^{-1} \left[ \mathcal{Q}_3 + \mathcal{H} \left( 2 \mathcal{Q}_1 - \mathcal{Q}_2 - \mathcal{Q}_4^{\prime} \right) \right]^{\prime}.
\end{eqnarray}
Finally, the solution is
\begin{eqnarray}
\psi^{(2)}_k & = & - \frac{1}{\tau^2} \left( \frac{\cos\zeta}{\zeta} + \sin\zeta \right) \int\limits \mathcal{Q}_{0k} \frac{\tau^3}{\zeta} \left( \frac{\sin\zeta}{\zeta} - \cos\zeta \right) d\tau + \\
& & + \frac{1}{\tau^2} \left( \frac{\sin\zeta}{\zeta} - \cos\zeta \right) \int\limits \mathcal{Q}_{0k} \frac{\tau^3}{\zeta} \left( \frac{\cos\zeta}{\zeta} + \sin\zeta \right) d\tau, \nonumber
\end{eqnarray}
with $\mathcal{Q}_{0k}$ denoting Fourier image of $\mathcal{Q}_0$.
Functions $\phi^{(2)}$ and $\sigma^{(2)}$ then can be found with the use of this solution and relations (\ref{eq:fipsi}) and (\ref{eq:sigma}).

\section{Summary and outlook}\label{sec:6}

We have studied solid matter filling a flat FLWR universe with shear stress Lam\'e  parameter, or shear modulus, set to zero.
Behaviour of such matter departs from the case with perfect fluid only in nonlinear perturbative regime.
In order to focus on the most physically interesting models we restricted ourselves to two cases, radiation-like solid with pressure to energy density ratio $w=1/3$, and dark energy-like solid with $w=-1$.
Additionally, we have imposed validity of condition for time-time component and space part trace of the mixed stress-energy tensor, $\widetilde{w}=-3T_{0}^{\phantom{0}0}/T_{i}^{\phantom{i}i}=$ const. (\ref{eq:condition}), in arbitrary coordinates, and restricted the parameter space of the solid matter model even more.

\vskip 1mm
From the technical point of view, when restricting the parameter space of the solid we have used only the condition (\ref{eq:condition}) assumed to be valid in arbitrary coordinates, because, as we have shown, it implies the above mentioned values of the pressure to energy density ratio.
This condition is inspired by the fact, that there are only two kinds of perfect fluid for which it is satisfied not only in its local rest frame, where $\widetilde{w}=w$ automatically, but also in arbitrary coordinates, namely radiation and dark energy.
In the case with radiation it is consequence of traceless stress-energy tensor in the Maxwell theory, $T_{\mu}^{\phantom{\mu}\mu}=0$, and in the case of dark energy it follows from proportionality of the stress-energy tensor to the space-time metric, $T_{\mu\nu}\propto g_{\mu\nu}$.

\vskip 1mm
The second order perturbative analysis of the condition (\ref{eq:condition}) assumed to be valid in arbitrary coordinates leads to only two allowed values of the pressure to energy density ratio, $w=1/3$ and $w=-1$, the same as for radiation and dark energy as perfect fluids.
Therefore, by passing from the perfect fluid to solid matter with zero shear modulus we do not obtain another physically interesting value of $w$, but there remains one additional parameter $\nu$ which differentiates the solid with zero shear modulus from the perfect fluid, and it can be arbitrary.
Coefficients $w$ and $\nu$ then parametrize the equation of state of the solid as in (\ref{eq:eqs00}).
For $\nu = - 2 (w+1) / 3$ the studied matter reduces to the special case coinciding up to the third order with the perfect fluid.

\vskip 1mm
In the second part of the paper, in sections \ref{sec:4} and \ref{sec:5}, we have checked that in the dark energy-like case with $w=-1$ no pure scalar perturbations up to second order can be formed, not only in the case of perfect fluid, but also for arbitrary value of the coefficient $\nu$.
With $\nu\neq 0$ the stress-energy tensor for the dark energy-like solid is no longer proportional to the space-time metric as in the perfect fluid case with $\nu=0$, although the condition (\ref{eq:condition}) remains valid in arbitrary coordinates.
Absence of scalar perturbations then had to be proven through analysis of equations for perturbations. 
In the same way, we have analysed scalar perturbations also in the radiation-like case with $w=1/3$.
In order to simplify the analysis we omitted the effect of other than scalar perturbations.
Such simplification is reasonable in our universe, as long as we are not interested in equations for second order vector and scalar perturbations which are sourced by first order perturbations.
The reason is that the first order vector perturbations decay and observations suggest small tensor to scalar ratio.
Consequently, the main contribution to observable effects comes from the scalar perturbations.

\vskip 1mm
Concepts studied in this paper including both dark energy-like and radiation-like solids can be applied to cosmic inflation and reheating which follows it. To include the dark energy-like solid into the inflationary context, one has to slightly depart from the value of pressure to energy density ratio $w=-1$ by the slow-roll parameter, $w=-1+2\epsilon/3$, where $\epsilon=-\dot{H}/H^2$. This work has been already done with the model of solid inflation \cite{gruzinov,endlich} and in works following it \cite{akhshik,bartolo}. After the end of inflation, during reheating, energy of inflaton fields is transferred to the radiation. It is in principle possible that matter with properties similar to solid matter studied in our work can be either formed during the reheating, or it can be a remnant of the solid inflation or its generalizations. The model of radiation-like case of the solid then could be partially applicable towards the end of reheating and beginning of the radiation dominated era.

\vskip 1mm
The model analysed in this paper can be studied further.
This may include:
\begin{itemize}
\item Investigating effect of linear order tensor perturbations, which are formed also in the dark energy-like case, on evolution of second order perturbations;
\item relaxing the validity of condition (\ref{eq:condition}) in arbitrary coordinates, which have led to only two values of pressure to energy density ratio ( i.e. studying cases with arbitrary $w$);
\item studying even higher orders of the perturbation theory, where one has to define more coefficients parametrizing the equation of state of the solid, and investigate constraints on them;
\item considering more general space-time metrics including also space-times beyond a cosmological context, which may require also a non-perturbative approach.
\end{itemize}
We leave this for future work.

\section*{Acknowledgements}
I would like to thank Vladim\'ir Balek for discussions, and pointing out a mistake in a formula in the previous version of the paper (relations (\ref{eq:mistake}) and (\ref{eq:mistakf}) in this version).

\appendix
\renewcommand{\thesection}{\Alph{section}}

\section{Mixed stress-energy tensor of perfect fluid}\label{app:a}

By definition a perfect fluid is isotropic in its local rest frame.
It is then fully specified by the equation of state,
\begin{eqnarray}
p = w \rho,
\end{eqnarray}
which relates its local rest frame pressure $p$ to its local rest frame energy density $\rho$.
Stress-energy tensor of a perfect fluid is given by relation
\begin{eqnarray}
T_{\mu\nu} = \left(\rho+p\right)u_{\mu}u_{\nu}+pg_{\mu\nu},
\end{eqnarray}
where $u^{\mu}=dx^{\mu}/d\tau$ is 4-velocity of the fluid elements with $\tau$ denoting their proper time.
Due to normalization, $u_{\mu}u^{\mu}=-1$, its zero component $u^0$ depends on its space part $u^i$, which is given by the space velocity $v^i$,
\begin{eqnarray}
u^i = \frac{dx^i}{dx^0}\frac{dx^0}{d\tau} = v^i u^0.
\end{eqnarray}
Thus, we can write
\begin{eqnarray}
\label{eq:mistake}
u^0=\frac{1}{\sqrt{-g_{00}-2g_{0i}v^i-g_{ij}v^iv^j}}.
\end{eqnarray}
The time-time component and the space part trace of the mixed stress-energy tensor in arbitrary coordinates can be written as
\begin{eqnarray}
\label{eq:a1}
T_{0}^{\phantom{0}0} & = & \rho \left[ -1 + \left(w+1\right) \xi \right], \\
T_{i}^{\phantom{i}i} & = & \rho \left[ 3w - \left(w+1\right) \xi \right], \nonumber  
\end{eqnarray}
where $\xi=-u_iu^i=1+u_0u^0$, which can be expressed as
\begin{eqnarray}
\label{eq:mistakf}
\xi = \frac{g_{0i}v^i+g_{ij}v^iv^j}{g_{00}+2g_{0i}v^i+g_{ij}v^iv^j}.
\end{eqnarray}
Now we can check facts about $T_{0}^{\phantom{0}0}$ to $T_{i}^{\phantom{i}i}$ ratio stated in section \ref{sec:1}.
For the radiation with $w=1/3$ relations in (\ref{eq:a1}) imply $T_{0}^{\phantom{0}0}=-T_{i}^{\phantom{i}i}=\rho(-1+4\xi/3)$, and for the dark energy with $w=-1$ we have $T_{0}^{\phantom{0}0}=T_{i}^{\phantom{i}i}/3=-\rho$.
In both cases the condition (\ref{eq:condition}) is satisfied in arbitrary coordinates, while for other choices of pressure to energy density ratio relations in (\ref{eq:a1}) imply that it does not.
This is obvious from expression of the quantity $\widetilde{w}$ defined in (\ref{eq:condition}) through quantities introduced here,
\begin{eqnarray}
\widetilde{w} = \frac{3w - \left(w+1\right) \xi}{3 - 3\left(w+1\right) \xi}.
\end{eqnarray}
If $w\neq1/3$ and $w\neq-1$ the relation (\ref{eq:condition}) holds only in the local rest frame of the fluid, in which $\xi=0$, and $\widetilde{w}=w$ automatically.
For example in the case with dust, $w=0$, and
\begin{eqnarray}
\left.\frac{T_{0}^{\phantom{0}0}}{T_{i}^{\phantom{i}i}}\right|_{w=0}=\frac{1}{\xi}-1\neq\textrm{const}.
\end{eqnarray}
In conclusion, the condition (\ref{eq:condition}) is satisfied for any perfect fluid with arbitrary pressure to energy density ratio, but only in its local rest frame, while validity of this condition in any coordinates is related to only two cases, $w=1/3$ and $w=-1$.

{\setstretch{1.0}

}

\end{document}